\magnification=1200
\hsize=15truecm
\vsize=23truecm
\baselineskip 20 truept
\voffset=-0.5truecm
\parindent=0cm
\overfullrule=0pt
\def\F{{\rm F}}
\def\R{{\rm R}}
\font\titolo=cmbx10 scaled\magstep2
\def\V{\hbox{\hbox{${V}$}}\kern-1.9mm{\hbox{${/}$}}}
\catcode`@=11
%
%
%
\def\lsim{\mathchoice
  {\mathrel{\lower.8ex\hbox{$\displaystyle\buildrel<\over\sim$}}}
  {\mathrel{\lower.8ex\hbox{$\textstyle\buildrel<\over\sim$}}}
  {\mathrel{\lower.8ex\hbox{$\scriptstyle\buildrel<\over\sim$}}}
  {\mathrel{\lower.8ex\hbox{$\scriptscriptstyle\buildrel<\over\sim$}}} }
\def\gsim{\mathchoice
  {\mathrel{\lower.8ex\hbox{$\displaystyle\buildrel>\over\sim$}}}
  {\mathrel{\lower.8ex\hbox{$\textstyle\buildrel>\over\sim$}}}
  {\mathrel{\lower.8ex\hbox{$\scriptstyle\buildrel>\over\sim$}}}
  {\mathrel{\lower.8ex\hbox{$\scriptscriptstyle\buildrel>\over\sim$}}} }
\def\croce{\displaystyle / \kern-0.2truecm\hbox{$\backslash$}}
\def\lqua{\lower4pt\hbox{\kern5pt\hbox{$\sim$}}\raise1pt
\hbox{\kern-8pt\hbox{$<$}}~}
\def\gqua{\lower4pt\hbox{\kern5pt\hbox{$\sim$}}\raise1pt
\hbox{\kern-8pt\hbox{$>$}}~}
\def\mma{\lower1pt\hbox{\kern5pt\hbox{$\scriptstyle <$}}\raise2pt
\hbox{\kern-7pt\hbox{$\scriptstyle >$}}~}
\def\mmb{\lower1pt\hbox{\kern5pt\hbox{$\scriptstyle >$}}\raise2pt
\hbox{\kern-7pt\hbox{$\scriptstyle <$}}~}
\def\mmc{\lower4pt\hbox{\kern5pt\hbox{$<$}}\raise1pt
\hbox{\kern-8pt\hbox{$>$}}~}
\def\mmd{\lower4pt\hbox{\kern5pt\hbox{$>$}}\raise1pt
\hbox{\kern-8pt\hbox{$<$}}~}
\def\lsu{\raise4pt\hbox{\kern5pt\hbox{$\sim$}}\lower1pt
\hbox{\kern-8pt\hbox{$<$}}~}
\def\gsu{\raise4pt\hbox{\kern5pt\hbox{$\sim$}}\lower1pt
\hbox{\kern-8pt\hbox{$>$}}~}
\def\croce{\displaystyle / \kern-0.2truecm\hbox{$\backslash$}}
\def\ali{\hbox{A \kern-.9em\raise1.7ex\hbox{$\scriptstyle \circ$}}}
\def\2frecce{\hbox{\lower 0.3ex\hbox{$\leftarrow$} 
\hbox{\kern-1.3em\raise 0.3ex\hbox{$\rightarrow$}}}}
%
%
%
%
\def\quad@rato#1#2{{\vcenter{\vbox{
        \hrule height#2pt
        \hbox{\vrule width#2pt height#1pt \kern#1pt \vrule width#2pt}
        \hrule height#2pt} }}}
\def\quadratello{\mathchoice
\quad@rato5{.5}\quad@rato5{.5}\quad@rato{3.5}{.35}\quad@rato{2.5}{.25} }
%
%
\font\s@=cmss10\font\s@b=cmbx8
\def\reali{{\hbox{\s@ l\kern-.5mm R}}}
\def\m{{\hbox{\s@ l\kern-.5mm M}}}
\def\k{{\hbox{\s@ l\kern-.5mm K}}}
\def\naturali{{\hbox{\s@ l\kern-.5mm N}}}
\def\interi{{\mathchoice
 {\hbox{\s@ Z\kern-1.5mm Z}}
 {\hbox{\s@ Z\kern-1.5mm Z}}
 {\hbox{{\s@b Z\kern-1.2mm Z}}}
 {\hbox{{\s@b Z\kern-1.2mm Z}}}  }}
\def\complessi{{\hbox{\s@ C\kern-1.7mm\raise.4mm\hbox{\s@b l}\kern.8mm}}}
\def\toro{{\hbox{\s@ T\kern-1.9mm T}}}
\def\unity{{\hbox{\s@ 1\kern-.8mm l}}}
%
%
\font\bold@mit=cmmi10
\def\setbmit{\textfont1=\bold@mit}
\def\bmit#1{\hbox{\textfont1=\bold@mit$#1$}}
%
\catcode`@=12

\null
\vskip 0.5truecm
\rightline{DFPD/95/TH/63}
\rightline{November 1995}
\vskip 1truecm

\centerline{\titolo The cancellation of worldsheet anomalies in the}

\centerline{\titolo D=10 Green--Schwarz heterotic string sigma--model${}^*$}

\vskip 2.0truecm
\centerline{\bf K. Lechner and M. Tonin}

\vskip 0.5truecm

\centerline{\it Dipartimento di Fisica, Universit\`a di Padova}
\centerline{\it and}
\centerline{\it Istituto Nazionale di Fisica Nucleare, Sezione di Padova}
\centerline{\it Italy}

\vskip 2truecm
\midinsert
\baselineskip 15truept
\centerline{\bf Abstract}
\vskip 0.3truecm
We determine the two--dimensional Weyl, Lorentz and $\kappa$--anomalies in 
the $D=10$ Green--Schwarz heterotic string sigma--model, 
in an $SO(1,9)$-Lorentz covariant background gauge,
and prove their cancellation.

\endinsert

\vskip 4truecm
$^*$ Supported in part by M.P.I. This work is carried out in the 
framework of the European Community Programme ``Gauge Theories, Applied 
Supersymmetry and Quantum Gravity" with a financial contribution under 
contract SC1--CT--92--D789.

\vfill\eject

\noindent
{\bf 1. Introduction}
\vskip0.5truecm

\qquad
The principal advantage of the Green--Schwarz string, with respect to 
the Neveu--Schwarz--Ramond formulation, is its manifest 
target space--time supersymmetry; its principal disadvantage is 
caused by the fundamental $\kappa$--symmetry 
which, being infinitely reducible, gives rise to problems with the
Lorentz--covariant gauge--fixing of this symmetry.
This last feature makes it 
rather difficult to compute the full worldsheet anomalies of the 
Green--Schwarz string, 
and the related sigma--model, and to prove finally their cancellation
in $D=10$.
To do that, most papers, refs. [1,2], used a non covariant, 
semi--light cone gauge. However, a direct calculation of the 
contribution of the fermionic string fields $\vartheta$ to the
Weyl anomaly in this gauge leads to a result which is 1/4 of the
correct value [1]. Possible ways to overcome this difficulty have been 
proposed in [2]. 

\qquad
To our knowledge there is actually only one paper, ref.[3], by P.B. 
Wiegmann, in which the conformal (Weyl) anomaly in the Green--Schwarz 
heterotic string has been determined in a ``covariant semi--light cone gauge", 
which is accessible in an on--shell configuration of the string, and shown 
to vanish in ten dimensions.\footnote*{The correct result is also obtained 
in the framework of the problematic Lorentz covariant gauge fixing 
involving an infinite tower of ghosts [4].}

\qquad
The present paper can be viewed as an extension of the technique
used in [3] in the following directions: first of all we determine the
worldsheet anomalies in the heterotic string Green--Schwarz 
{\it sigma--model}, using the background 
field method combined with a normal coordinate
expansion [5, 6]. This permits us to keep $SO(1,9)$ covariance manifest. The 
splitting of the string variables in classical and quantum fields, the 
classical fields being on shell, makes the covariant semi--light cone gauge
accessible.
Moreover, we determine the complete (Weyl, Lorentz and 
$\kappa)$--worldsheet anomaly due to the string coordinates $(X, \vartheta)$ 
and to the ghost ($b, c)$--system, and show that it cancels against 
the anomaly due to the 32 heterotic fermions.
Finally, our procedure establishes a deep connection existing between the 
worldsheet anomaly and the target space $SO(1,9)$ Lorentz--anomaly, whose 
understanding is, actually, crucial for the cancellation of the total 
worldsheet anomaly.

\qquad
Another important feature of our procedure is that it can be 
extended to other $p$--brane $\sigma$--models. In particular in [7]
this method is applied to compute the worldvolume anomalies
of the super fivebrane sigma--model, which is supposed to be dual 
to the $D=10$ heterotic string, [8].

\vskip0.7truecm
\noindent
{\bf 2. The action and the gauge fixing}
\vskip0.5truecm

\qquad
The sigma--model action for the heterotic Green--Schwarz string in ten 
target space--time dimensions is given by
$$
I = - {1 \over 2\pi \alpha'} \int d^2 \sigma \left({1\over2}
\sqrt{g}\ g^{i j} V_i^a V_{j a}
+{1\over2}\varepsilon^{i j} V^A_i V^B_j B_{BA} - {1\over2}\sqrt{g}\ e_-^j \psi 
(\partial_j - A_j) \psi \right). \eqno(1)
$$
The string fields are the supercoordinates $Z^M = (X^m (\sigma), 
\vartheta^\mu (\sigma))$, the 32 heterotic fermions $\psi (\sigma)$ and the
worldsheet zweibeins $e_{\pm}^i (\sigma), g^{ij} = e_+^{(i} e_-^{j)}$.
The induced zehnbeins are 
given by $V_i^A = \partial_i Z^M E_M{}^A (Z)$ and the $SO(1,9)$ flat 
index $A=(a, \alpha)$ stands for ten bosonic ($a=0,...,9)$ and 
sixteen fermionic $(\alpha =1, ..., 16)$ entries. The induced target space 
connections are $\Omega_{i a}{}^b (Z) = V_i^C \Omega_{Ca}{}^b$ for 
$SO(1,9)$, and $A_i(Z) = V_i^{C} A_C$ for the gauge group
$SO(32)$. Flat two--dimensional light--cone indices for a worldsheet vector
$W_i$ are introduced via $W_{\pm} = e_{\pm}^i W_i, V^A_{\pm} = e_{\pm}^{i} 
V_i^{A}, \partial_{\pm} = e_{\pm}^{i} \partial_i$ etc. (see ref. [6] for the 
notation). Two--dimensional flat {\it vector} indices are indicated by an 
index with a ``hat", $\widehat a$, e.g. $e_{\widehat a}^i \ (\widehat a 
=0,1)$. The action (1) is invariant under the transformations
$$
\eqalign{
\delta Z^M = & \Delta^\alpha E_\alpha{}^M + c^i \partial_i Z^M
\cr
\delta \psi = & \left({1\over 2} (\ell + \lambda) + \Delta^\alpha A_\alpha + 
c^i \partial_i \right) \psi \cr
\delta e_+^i = & (\lambda + \ell) e_+^i - 4  e_-^{i} 
\left(V_+^\alpha -{1\over 2} 
\psi \chi^\alpha \psi\right) \kappa_\alpha + c^j \partial_j e_+^i - 
\partial_+ c^i \cr
\delta e_-^i = & (\lambda - \ell) e_-^i + c^j \partial_j e_-^i - 
\partial_- c^i. \cr} \eqno(2)
$$
We indicate the ghost fields for worldsheet
Weyl, Lorentz, diffeomorphism and
$\kappa$--transformations respectively with $\lambda, \ell, c^i$ and 
$\kappa_\alpha \equiv \kappa_{+ \alpha}$; $\Delta^\alpha = 
(\Gamma^a)^{\alpha\beta} V_{-a} \kappa_\beta$. $\chi^\alpha(Z)$ is the
ten dimensional gluino superfield. 
Invariance under 
$\kappa$--transformations is, actually, achieved if the target space fields
satisfy suitable superspace constraints, given in the Appendix.
There we report also the BRST transformations of the ghost fields which, 
together with (2), give rise to a BRST operator $\Omega$ which closes if 
the string fields satisfy their equations of motion: $\Omega^2 = 0$ 
(on--shell). Having such an operator is extremely useful in that it
allows to determine the $\kappa$--anomalies, once the Weyl and 
$SO(1,1)$--anomalies are known, upon enforcing the Wess--Zumino consistency 
condition (see below).

\qquad
To compute the total anomaly we proceed as follows. We use the background 
field method along with a normal coordinate expansion [5,6], writing
$Z = Z_0 + \Pi (Z_0, y)$ and  $\psi = \psi_0 + \psi_q$, where $(Z^M_0 ,
\psi_0$) are external ``classical" fields, and $\psi_q$ and the normal 
coordinates $y^A = (y^a, y^\alpha)$ are ``quantum" fields over which we 
perform the functional integration. The functions $\Pi^M$ trigger the 
manifest $SO(1,9)$ Lorentz covariance of the background field method. 
Since we choose to maintain the effective action diffeomorphism invariant, 
at the expense of local worldsheet Weyl--Lorentz anomalies, the ghost fields 
$c^i$ (and the antighosts $b_{ij}$) can be treated as purely 
``quantum"; the zweibeins  $e_{\pm}^i$ as well as the ghosts $\ell,\lambda,
\kappa$
are considered as purely ``classical". The classical fields transform
according (the classical counterparts of) eq. (2). Moreover from now on, the 
pullback zehnbeins $V_i^A$, the Lorentz connection $\Omega_{iab}$, 
the gauge connection $A_i$ etc. will be the classical ones, i.e. will be 
evaluated at $Z=Z_0$. Finally, we set the classical fields on shell.

\qquad
Since the action is invariant under (2) and $\Omega^2 =0$ on--shell, even if
the heterotic fermions are absent, we can derive first the anomaly ${\cal 
A}_1$, gotten by the functional integration over $(y^A, b,c)$ for $\psi_0 
=0$. The dependence on $\psi_0$ of this anomaly can be retrieved by 
enforcing the Wess--Zumino consistency condition on ${\cal A}_1, \Omega {\cal
A}_1 =0$. Then we perform the functional integration over $\psi_q$, 
compute the related anomaly ${\cal A}_2$ and show that ${\cal A}_1 + {\cal 
A}_2 =0$.

\qquad
The core of the present paper is constituted by the $SO(1,9)$--covariant 
functional integration over the fermionic $y^\alpha$ which we perform 
below.
\qquad

For $\psi_0=0$ the equations of motion of the metric in terms of classical 
fields becomes
$$
V^a_i V_{j a} = {1 \over 2} g_{ij}  V^a_+ V_{a-} \equiv e^{2 \phi} 
g_{ij}. \eqno(3)
$$
The use of eq. (3) allows to perform a Lorentz--covariant $\kappa$--gauge 
fixing on the $y^\alpha$. We define the matrix
$$
\Gamma^\alpha{}_\beta = {1\over 2} e^{-2 \phi} {\varepsilon^{i j} \over 
\sqrt{g}}  V^a_i V^b_j (\Gamma_{ab})^\alpha{}_\beta
$$
which, due to (3), satisfies $\Gamma^2= \unity$,  $tr \Gamma =0$. 
Since the $\kappa$--transformation law for $y^\alpha$ is 
$\delta_\kappa y=\V_-\kappa_q+o(y)$, where $\kappa_q$ is the
quantum ghost, the condition
$$
{\unity + \Gamma \over 2} y= 
{e^{-2\phi}\over 4} \V_- \V_+ y=0 \eqno(4)
$$
eliminates just 8 of the 16 $y's$ and fixes $\kappa$--symmetry.
Moreover, in the gauge (4), being algebraic, the ghost--fields 
$\kappa_q$ do not propagate.

\qquad
A second essential ingredient we need is the knowledge of the target space 
$SO(1,9)$ Lorentz anomaly of the effective action, which is due to the non 
invariance of the integration measure $\int \{{\cal D} y\}$ under local 
$SO(1,9)$ transformations.

\qquad
In a non--covariant gauge this anomaly has been computed in ref. [6]; the 
techniques used  there can be adapted for the gauge (4) and the result is 
\footnote*{The non trivial part of the anomaly is clearly independent on the 
gauge--fixing.}
$$
{\cal A} (L) = {1 \over 8 \pi} \int d^2 \sigma \sqrt{g}\ tr 
(\partial_- L \widetilde\Omega_+). \eqno(5)
$$
$L \equiv L_{ab}$ is the infinitesimal
$SO(1,9)$ transformation parameter and $\widetilde 
\Omega_+{}^{ab}$ is defined by 
$$
\widetilde \Omega_+{}^{ab} = \Omega_+{}^{ab} - e^{-2 \phi} V_+^{[a} D_+ 
V_-^{b]} - T_+{}^{ab}, \eqno(6)
$$
$$
D_{\pm} = \partial_{\pm} \pm \omega_{\pm} + \Omega_{\pm *}{}^*, \eqno(7)
$$
where $\omega_{\pm}$ are the worldsheet connections
$$
\omega_{\pm} = \pm {1 \over \sqrt{g}}\ \partial_j (\sqrt{g}\ e_{\pm}^{j}), 
\eqno(8)
$$
in terms of which the scalar curvature becomes ${\cal R}^{(0)}= D_- \omega_+ - 
D_+\omega_-$.

\qquad
The explicit expression of $T_+{}^{ab}$ is given in the Appendix, here it 
suffices to know that
$$
T_{+ab} V_-^{b} =0\ . \eqno(9)
$$
With the symbol $tr$ we indicate 
the trace in the vector representation of $SO(1,9)$ or 
of $SO(1,1)$ since no confusion should arise.

\qquad
Under {\it finite} $SO(1,9)$ transformations, with transformation parameter 
$\Lambda^a{}_b$, the measure $\{{\cal D}y\}$ 
changes, according to (5), by a Wess--Zumino action given by
$$
\Gamma_{WZ} (\Lambda) = {1\over 8 \pi} \left(\int d^2 \sigma \sqrt{g}\ tr
\left(\Lambda^T \partial_- \Lambda \widetilde \Omega_+ - {1\over 2} g^{ij} 
\partial_i \Lambda^T \partial_j \Lambda \right) - {1 \over 3} \int_{D_3} tr
\left(d \Lambda \Lambda^T\right)^3 \right), \eqno(10)
$$
where $\Lambda^a{}_b$ satisfies $\Lambda^a{}_b \Lambda^c{}_d \eta^{bd} = 
\eta^{ac}$. $D_3$ is a three--dimensional manifold with the string
worldsheet as boundary.

\vskip0.7truecm
\noindent
{\bf 3. The anomaly computation}
\vskip0.5truecm

\qquad
Now we are able to perform the functional integration over the $y^\alpha$. 
After normal coordinate expansion the relevant contribution to the expanded
action is given by the $SO(1,9)$--invariant kinetic term for the $y^\alpha$
$$
{1 \over 4} \int d^2 \sigma \left(\sqrt{g}\ g^{ij} + \varepsilon^{ij}\right)
\ y\ \Gamma_a V^a_i D_j y
$$
which, upon enforcing (4), can be written as
$$
I (V, \Omega, y) = {1 \over 2} \int d^2 \sigma \sqrt{g}\ g^{ij} V^a_i \
y\ \Gamma_a 
{\unity - \Gamma \over 2} D_j {\unity - \Gamma \over 2} \ y \eqno(11)
$$
where $D_j \equiv \partial_j - {1 \over 4} \Gamma{}_{cd} 
\Omega_j{}^{cd}$.

\qquad
The normal--coordinate--expanded action contains actually additional terms 
quadratic in the $y^\alpha$; for what concerns the $SO(1,9)$ anomaly (5) 
these terms give just rise to the contribution proportional to $T_{+ab}$ in
eq. (5), while they do not contribute to the worldsheet Lorentz 
and Weyl anomalies. 

\qquad
The principal problem related with (11) is 
that its kinetic term is not canonical in the sense that it is multiplied 
by the external field $V_i^a$ which is not constant.
On the other hand (11) is $SO(1,9)$ Lorentz 
invariant and this invariance can be used to eliminate this
unwanted dependence
on $V_i^a$. For a generic $SO(1,9)$ transformation $\Lambda^a{}_b$ we have 
$I(V, \Omega, y)= I (V^\Lambda, \Omega^\Lambda, y^\Lambda)$, and changing 
integration variable from $y$ to $y^\Lambda$ we can replace $I(V, \Omega, 
y)$ with $I(V^\Lambda, \Omega^\Lambda, y)$.
But, since the measure $\int \{{\cal D} y\}$ is not invariant, this change
of variable results in the appearance of $\Gamma_{WZ} (\Lambda)$ as given 
in (10).
Therefore the integration over the fermionic $y^\alpha$ results in an effective 
action, $\Gamma_F$, which can be written as 
$$
\Gamma_F = \Gamma_0 - \Gamma_{WZ} (\Lambda) \eqno(12)
$$
where
$$
e^{i \Gamma_0}= \int \{{\cal D}y\} e^{iI(V^\Lambda, \Omega^\Lambda, y)}\ 
. \eqno(13)
$$
\qquad
It remains to choose an appropriate $\Lambda^a{}_b$. For this purpose
we introduce eight $SO(1,9)$ Lorentz vectors, $N_a^r\ (r=2,...,9)$ satisfying
$$\eqalign{
N^r_a N^{as} = & -\delta^{rs} \cr
N^r_a  V^a_j =& 0 \cr} \eqno(14)
$$
and choose
$$\eqalign{
\Lambda^{\widehat a}{}_a = & e^{-\phi} \widetilde e^{j\ \widehat a}
V_{ja} \cr
\Lambda^r{}_a = & N^r_a\ (r=2,...,9) \cr
\widetilde e^j_\pm = & e^{\pm \phi} e^j_\pm\ . \cr} \eqno(15)
$$
Due to (3), (14) and $\widetilde e{}_{\widehat a}{}^i \widetilde 
e{}_{\widehat 
b}{}^j g_{ij} = \eta_{\widehat a\ \widehat b}$, this $\Lambda$ is indeed an 
element  of $SO(1,9)$, in that $\Lambda^a{}_b \Lambda^c{}_d\eta^{bd} 
= \eta^{ac}$.
With this choice the kinetic term for $y^\alpha$ becomes indeed canonical
$$
I(V^\Lambda, \Omega^\Lambda, y) = {1 \over 4} \int d^2
\sigma\sqrt{g}\ e^j_+\ y\
\Gamma_- \Bigl(\partial_j - {1 \over 4} \Gamma_{rs} 
W_j{}^{rs}\Bigl)y\ , \eqno(16)
$$
where $\Gamma_- \equiv (\Gamma_0 - \Gamma_1)$ is a constant matrix and 
projects out just eight of the sixteen $y's$, and $W_j{}^{rs} 
\equiv N_a^s
(\partial_j N^{ar} - \Omega_j{}^{ab} N^r_b)$ is Weyl, $SO(1,1)$ and $SO(1,9)$
{\it invariant} and does, therefore, not affect the corresponding 
anomalies. The relevant contribution to $\Gamma_0$ is therefore just given 
by 8 $\ell n^{1/2} \det (\sqrt{g}\ \partial_+)$ which is equal to
$$
\Gamma_0 = {1 \over 48\pi} \int d^2\sigma\sqrt{g}\ tr 
\Bigl(D_- \omega_+\ {1\over 
\quadratello}\ D_- \omega_+\Bigr)\ . \eqno(17)
$$
Here we defined $\omega_{\pm\widehat a\ \widehat b} = \omega_{\pm}
\varepsilon_{\widehat a\ \widehat b}$. 
Under worldsheet Lorentz and Weyl transformations we have
$$
\delta \Gamma_0 = -{1 \over 24\pi} \int d^2\sigma\sqrt{g}\ tr 
\left(\partial_- 
(\ell - \lambda) \omega_+\right) \eqno(18)
$$
$$
- \delta \Gamma_{WZ} = -{1 \over 8\pi} \int d^2 \sigma 
\sqrt{g}\ tr \left(\partial_- 
(\ell - \lambda) \omega_+\right)\ , \eqno(19)
$$
where we wrote $\lambda_{\widehat a\ \widehat b}=\varepsilon_{\widehat a\
\widehat b}\ \lambda, \ell_{\widehat a\ \widehat b}=\varepsilon_{\widehat a\
\widehat b}\ \ell$.
The evaluation of $\delta\Gamma_{WZ}$
is long but straightforward. One has to use (6) together with (9), and the
decomposition 
$$
\omega_{+\widehat a \ \widehat b} = e^{-2\phi} \left( 
V_{[\widehat b}^a\partial_+ V_{\widehat a]a}-\Omega_{+ab}
V_{\widehat b}^a V_{\widehat a}^b\right) +
\left( {1\over 2} e^{-2\phi} D_+ V_-^a V_{+a} - \partial_+\phi
\right) \varepsilon_{\widehat a\  \widehat b}
$$
which follows from the embedding equation (3).

\qquad
We see that the effect 
of the Wess--Zumino term is just to quadruplicate the ``naif" result,
$\delta \Gamma_0$, which corresponds to the Weyl--Lorentz anomaly of just 
eight quantum $\vartheta ' s$. The contribution of the Wess--Zumino term, 
which is actually essential for the cancellation of the total anomaly, is 
missed in non--covariant perturbative approaches, refs. [1].

\qquad
The relevant contribution to the effective action of the bosonic 
coordinates and the ghost fields is standard $(D=10)$ 
$$
\Gamma_{y^a,b,c} = {D-26\over 96 \pi} \int d^2\sigma\sqrt{g}\ {\cal R}^{(0)}\ 
{1 \over 
\quadratello}\ {\cal R}^{(0)}\ , \eqno(20)
$$
whose variation under Weyl and $SO(1,1)$ is just 
$$
\delta\Gamma_{y^a,b,c} = {1 \over 3\pi} \int d^2\sigma\sqrt{g}\ tr 
\left((D_- \omega_+ - D_+ \omega_-) \lambda \right)\ .
$$
Summing up $\delta\Gamma_{y^a,b,c}$, $\delta \Gamma_F$ and the variation of 
the local term ${1 \over 6\pi} \int d^2\sigma\sqrt{g}\ tr (\omega_+\omega_-)$ 
one gets for the total
Weyl--Lorentz anomaly due to the fields $(y^\alpha,y^a,b,c)$
$$
{\cal A}_{\lambda,\ell} = {1 \over 6\pi} 
\int d^2\sigma\sqrt{g}\ tr \left(\partial_+ (\ell
+ \lambda) \omega_-\right)\ . \eqno(21)
$$
The corresponding $\kappa$--anomaly can be determined by enforcing the 
consistency condition on the complete anomaly, ${\cal A} = {\cal 
A}_{\lambda, \ell} + {\cal A}_\kappa,\ \Omega {\cal A} = 0$. This determines
${\cal A}$ as
$$
{\cal A} = {1 \over 6 \pi} \int d^2\sigma\sqrt{g}\ tr 
\left( \partial_+ (\ell + 
\lambda) \omega_- -2 \omega_- \omega_- 
V_+^\alpha \kappa_\alpha\right)\ . \eqno(22)
$$
So far we have set the heterotic fermions to zero. If they are present, eq.
(3) gets modified to
$$
V_i^a V_{ja} = g_{ij} e^{2\phi} + {1 \over 4} e_{-i} e_{-j} 
\psi_0 (\partial_+ - A_+) \psi_0\ , \eqno(23)
$$
where $\psi_0$ are the classical heterotic fermions, and in this case 
$\Lambda$, as given in (15), does no longer belong to $SO(1,9)$. However, by 
defining a modified metric, $g^*_{ij}$, through
$$\eqalign{
e^*_{+j} = &  e_{+j} + {1\over 4}\ e^{-2 \phi} 
e_{-j} \psi_0 
(\partial_+ - A_+)\psi_0 \cr
e^*_{-j} = & e_{-j}\ , \cr} \eqno(24)
$$
one can rewrite (23) as $V^a_i V_{ja} = g^*_{ij} e^{2\phi}$ and, 
using this, one can again construct a $\Lambda^*\in\ SO(1,9)$ along 
the same lines which brought to (15). The shift in (24) cannot modify the 
Weyl--Lorentz anomaly, (21), gotten by the functional integration over 
$(y^\alpha, y^a,b,c)$, but only the $\kappa$--partner, by a term quadratic in 
$\psi_0$. We can again enforce the Wess--Zumino consistency condition on 
${\cal A}_1 = {\cal A} + {\cal A}'_\kappa,\ \Omega\ {\cal A}_1 =0$,
which gives now:
$$
{\cal A}_1 = {1 \over 6\pi} \int d^2\sigma\sqrt{g}\ tr 
\left(\partial_+ (\ell + 
\lambda) \omega_- - 2 \omega_- \omega_- 
\left(V_+{}^\alpha - {1\over 2}\ \psi_0 
\chi^\alpha \psi_0\right)\kappa_\alpha \right)\ . \eqno(25)
$$
\qquad
The contribution of the 32 quantum heterotic fermions to 
the effective action (for 
$A_-=0$) is standard and corresponds to
$$
\Gamma_\psi = 32\ \ell n^{1/2}\ \det\ \left(\sqrt{g}\ \partial_-\right)= 
{1\over 12\pi}
\int d^2 \sigma \sqrt{g}\ tr \Bigl(D_+\omega_- {1\over\quadratello} D_+ \omega_-
\Bigr)\ .\eqno(26)
$$
Since we have the following total variations:
$$\eqalign{
\delta\omega_-& =\partial_- (\ell+\lambda) + (\lambda-\ell)\omega_-\cr
\delta \left(\sqrt{g}\ D_+ \omega_-\right) & = \sqrt{g}\ 
\left(\quadratello\ (\ell+\lambda) - 
4D_-\left(\omega_-\left(V_+{}^\alpha - {1\over 2}\ \psi_0 \chi^\alpha 
\psi_0\right)
\kappa_\alpha\right)\right)\cr
\quadratello & = D_+\partial_- = D_- \partial_+\ , \cr}
$$
one can easily compute ${\cal A}_2=\delta\Gamma_\psi$ and verify that
indeed
$$
{\cal A}_1 + {\cal A}_2 =0\ .
$$
\vskip0.7truecm
\noindent
{\bf 4. Some final remarks}
\vskip0.5truecm

\qquad
Our procedure reveals a connection between the 
$SO(1,1)$ and $SO(1,9)$ anomalies which emerges as follows.
The invariant polynomial corresponding to the target Lorentz
anomaly (5) is given by $X_4^L (\R) = {1\over 8\pi}tr(\R \R)$
where $\R$ is the $D=10$ Lorentz curvature two--form. On the other hand, 
the ``naif" contribution of the eight physical $y^\alpha$ to the 
$SO(1,1)$ anomaly, 
eq. (18), corresponds to the invariant polynomial $X_4^{(0)}({\cal R}) 
= - {1\over 
24\pi} tr ({\cal R}{\cal R})$, where ${\cal R}$ is the $d=2$ Lorentz 
curvature two--form. 
What we have shown is that the total $d=2$ anomaly polynomial, 
corresponding to the $y^{\alpha}$, is given by $X_4^y ({\cal R}) 
= X_4^{(0)} ({\cal R}) - 
X_4^L ({\cal R})= -{1\over 6\pi}\ tr({\cal R}{\cal R})$.
The contribution  to $X_4$ from the $\psi$ is just 
(see (26)) 
$X_4^\psi ({\cal R}) = {1\over 6\pi}\ tr({\cal R}{\cal R})$, 
and $X_4^\psi + X_4^y =0$.

\qquad
Our procedure for computing anomalies required the introduction of eight 
$SO(1,9)$ vectors, $N_a^r\ (r=2,...,9)$, which span the (eight--dimensional)
space orthogonal to the $V_j^a$. Classically these base vectors are defined
only up to an (extrinsic) local $SO(8)$ rotation, $\delta N_a^r=\ell^{rs} 
N_{as}$, where $\ell^{rs}=-\ell^{sr}$. For consistency the quantum 
effective action should depend only on the orthogonal space but not on the 
particular basis $\{N_a^r\}$ we have chosen. It can, actually, be verified
that neither $\Gamma_0$ nor $\Gamma_{WZ}$ are $SO(8)$ invariant, but that
$\Gamma_F=\Gamma_0-\Gamma_{WZ}$ is indeed invariant, as expected.

\qquad
Let us also notice that under target--space $SO(1,9)$ rotations we have 
$\delta_L(\Gamma_0-\Gamma_{WZ})=-\delta_L \Gamma_{WZ}= {1\over 8\pi}
\int d^2\sigma\sqrt g\ tr (\partial_-L\widetilde\Omega_+)$ which reproduces
correctly the anomaly (5).

\qquad
What we have considered in this paper are the ``genuine string" 
worldsheet anomalies in the heterotic string sigma--model, i.e. those
anomalies which survive (in non critical dimensions) even when the target
fields are switched off. The ``genuine sigma--model" anomalies at one loop,
which go to zero when the target fields go to zero, have been determined in 
ref. [6]. For completeness we recall the result:
$$
{\cal A}_\sigma = - {1\over 16\pi} \int d^2 \sigma\varepsilon^{ij} V_i^A V^B_j
\Delta^\gamma (\omega_{3YM} - \omega_{3L})_{\gamma BA}\ ,  \eqno(27)
$$
where $\omega_{3YM}$ and $\omega_{3L}$ are the Yang--Mills and 
$SO(1,9)$--Lorentz Chern--Simons three-superforms satisfying $d\omega_{3L}
=tr(\R\R)$, $d\omega_{3YM}=tr(\F\F)$.
${\cal A}_\sigma$, which is a pure
$\kappa$--anomaly, is cancelled by defining the generalized supercurvature
$$
H=dB + {\alpha'\over 4} (\omega_{3YM} - \omega_{3L}) \eqno(28) 
$$ 
and imposing on it, rather than on $dB$, the constraints $(IV,V)$ in the
appendix. The Bianchi identity associated to (28),
$$
dH={\alpha'\over 4}(tr(\F\F)-tr(\R\R)),
$$
can then be consistently solved in superspace at first order in
$\alpha'$, see refs.[6,9].
Eq. (28) implies also an anomalous transformation law for $B$, 
which is just the right one to cancel the $SO(1,9)$ and $SO(32)$ anomalies 
associated to the anomaly polynomial
$$
X_4={1\over 8\pi}(tr(\R\R)-tr(\F\F)).
$$

\vfill\eject
{\bf APPENDIX}

\vskip 0.5truecm

{\bf 1.} The action (1) is $\kappa$--invariant if the target space fields 
satisfy suitable constraints. If we define the target space 
super--differential $d=dZ^M {\partial\over\partial Z^M}$ and 
$E^A = dZ^M E_M{}^A$ and $T^A = dE^A + E^B \Omega_B{}^A \equiv {1\over 2} E^B
E^C T_{CB}{}^A, F=dA + AA \equiv \hfill\break
{1\over 2} E^A E^B F_{BA}, H=dB\equiv
{1\over 6} E^A E^B E^C H_{CBA}$ these constraints are given by

$$\eqalign{
T_{\alpha\beta}{}^a & = 2(\Gamma^a)_{\alpha\beta}\quad (I) \cr
T_{a\alpha}{}^b  & = 0 \quad\quad\quad (II)\cr
F_{\alpha\beta}  & = 0 \quad\quad\quad (III) \cr
H_{\alpha\beta\gamma} & = 0 = H_{ab\alpha}\quad (IV)\cr
H_{a\alpha\beta} & = 2 (\Gamma_a)_{\alpha\beta}. \quad (V)\cr}
$$

$(III)$ implies in particular that $F_{a\alpha} = 2 (\Gamma_a)_{\alpha\beta}
\chi^\beta$, where $\chi^\beta$ is the gluino, with values in the Lie 
algebra of $SO(32)$, and $(I,II)$ imply that $T_{\alpha\beta}{}^\gamma = 
2\delta^\gamma_{(\alpha}\lambda_{\beta )} - (\Gamma^a)_{\alpha\beta}
(\Gamma_a)^{\gamma\delta}\lambda_\delta$, where 
$\lambda_\alpha=D_\alpha\varphi$, $\varphi$ being
the dilaton superfield.

\vskip 0.3truecm

{\bf 2.} The BRST transformations for the ghost fields which, together with 
(2), give rise to an on--shell nihilpotent  BRST operator, $\Omega^2=0$, are
given by:

$$\eqalign{
\delta\ell & = - c^j\partial_j\ell - (\partial_-+\omega_-)(\kappa\V_- 
\kappa)\cr
\delta\lambda & = - c^j\partial_j\lambda + (\partial_- -\omega_-)
(\kappa\V_-\kappa)\cr
\delta c^i &= - c^j\partial_j c^i + 2e_-^i 
(\kappa\V_-\kappa)\cr
\delta\kappa_\alpha & = - c^j\partial_j \kappa_\alpha + (\lambda-\ell) 
\kappa_\alpha - (V_-^a\lambda_\alpha+(\Gamma^a)_{\alpha\gamma}
V_-^\gamma) (\kappa\Gamma_a\kappa)\cr
& + (4V_-^\gamma\kappa_\gamma - \kappa \V_-\lambda) \kappa_\alpha -
\Delta^\varepsilon \Omega_{\varepsilon\alpha}{}^\gamma \kappa_\gamma.\cr}
$$

\vskip 0.3truecm

{\bf 3.} The quantity $T_+{}^{ab}$ in (6) can be read off from eq. (56)
of ref. [6] and is given by 
$$
T_+{}^{ab} = \widetilde T^{abc} V_{+c} + e^{-2\phi}
\widetilde T^{[a}{}_{cd} V_+{}^c V_-{}^d V_+{}^{b]}
$$
where the completely antisymmetric tensor $\widetilde T^{abc}$ is given by
$$
\widetilde T^{abc} = T^{abc} + e^{-2\phi}
\left({1\over 2}
(\Gamma^{abc})_{\alpha\beta} V_+{}^\alpha V_-{}^\beta
-{1\over 16}(\Gamma_g\Gamma^{abc})_\gamma{}^\alpha V^g_-
V_+{}^\beta
T_{\alpha\beta}{}^\gamma\right),
$$
and $T^{abc}$ is the supertorsion.
\vfill\eject

{\bf References}
\vskip 0.3truecm

\item
{[1]} U. Kraemmer and A. Rebhan, Phys. Lett. {\bf B236}
(1990), 255; 
F. Bastianelli, P. Van Nieuwenhuizen and A.
Van Proeyen, Phys. Lett. {\bf B253} (1991), 67.

\vskip 0.3truecm

\item
{[2]} S. Carlip, Nucl. Phys. {\bf B284} (1987), 365; 

R. Kallosh and A. Morozov, Intern. J. Mod. Phys. {\bf A3} (1988), 1943;

A.H. Diaz and F. Toppan, Phys. Lett. {\bf B211} (1988) 285;

M. Porrati and P. Van Nieuwenhuizen, Phys. Lett. {\bf B273} (1991), 47.

\vskip 0.3truecm

\item
{[3]} P.B. Wiegmann, Nucl. Phys. {\bf B323} (1989) 330.

\vskip 0.3truecm

\item
{[4]} U. Lindstrom, M. R$\check o$cek, W. Siegel, P. Van Nieuwenhuizen and 
A. Van De Ven, Phys. Lett. {\bf B227} (1989) 87; {\bf 228} (1989), 53; 

M.B. Green and C. Hull, Phys. Lett. {\bf B225} (1989), 57;
{\bf B229} (1989) 215;

R. Kallosh, Phys. Lett. {\bf B224} (1989) 275; {\bf B225} (1989) 49; 

S. Gates, M. Grisaru, U. Lindstrom, M. R$\check o$cek, W. Siegel, P. Van 
Nieuwenhuizen and A. Van De Ven, Phys. Lett. {\bf B225} (1989) 44.

\vskip 0.3truecm

\item
{[5]} J. Honerkamp, Nucl. Phys. {\bf B36} (1072), 130;

S. Mukhi, Nucl. Phys. {\bf B264}, (1086), 640.

M.T. Grisaru and D. Zanon, Nucl. Phys. {\bf B310} (1988) 57.

\vskip 0.3truecm

\item
{[6]} A. Candiello, K. Lechner and M. Tonin, Nucl. Phys. {\bf B438} (1995), 
67.

\vskip 0.3truecm

\item
{[7]} K. Lechner and M. Tonin, DFPD/96/TH/07.

\vskip 0.3truecm

\item
{[8]} M. J. Duff, Classical quantum gravity, {\bf 5} (1990), 167.

\vskip 0.3truecm

\item
{[9]} K. Lechner, Phys. Lett. {\bf B357} (1995), 57.

\bye